\documentclass[11pt,reqno]{amsart}
\usepackage{amsmath}
\usepackage{array}
\newtheorem{theorem}{Theorem}

\newtheorem{lemma}[theorem]{Lemma}
\newtheorem{corollary}[theorem]{Corollary}

\newtheorem{remark}[theorem]{Remark}

\input{amssym.def}
\input{amssym}
\def\be{\begin{equation}}
\def\ee{\end{equation}}
\def\om{\omega}

\def\Sc{Schr\"o\-din\-ger}
\def\la{\langle}
\def\be{\begin{equation}}
\def\ee{\end{equation}}
\def\ra{\rangle}
\def\ds{\displaystyle}

\def\om{\omega}

\def\R{\Bbb R} 
\def\Z{\Bbb Z} 
\def\Na{\Bbb N}
\def\T{\Bbb T}  
\def\C{\Bbb C}

\def\Re{{\rm Re}\,}
\def\Im{{\rm Im}\,}
\def\Sc{Schr\"o\-dinger}

\def\B{\cal B}
\def\H{\cal H}
\def\G{\cal G}

\def\ep{\epsilon}

\def\im{{\rm Im}}

\def\Sc{Schr\"odinger}
\def\P{{\mathcal P}}
\def\PT{{\mathcal P}{\mathcal T}}
\def\l{{\lambda}}

\def\S{{\mathcal S}}

\def\Im{{\rm Im}}

\def\nl{\newline}
\def\pn{\par\noindent}
\def\L{{\mathcal L}}

\def\B{{\mathcal B}}

\def\H{{\mathcal H}}
\def\M{{\mathcal M}}
\def\V{{\mathcal V}}
\def\W{{\mathcal W}}

\def\G{{\mathcal G}}  
\def\F{{\mathcal F}} 
\def\F{{\mathcal F}}

\def\F{{\mathcal F}}

\def\v{{\mathcal \nu}}

\def\Na{\Bbb N}

\def\A{{\mathcal A}}
\def\M{{\mathcal M}}

\def\la{\langle}
\def\bea{\begin{eqnarray}}
\def\eea{\end{eqnarray}}
\def\ra{\rangle}
\def\ds{\displaystyle}
\def\om{\omega}

\def\ep{\varepsilon}

\begin{document}
\baselineskip=20pt
\title{Convergent Quantum Normal Forms, ${\mathcal P}{\mathcal T}$-symmetry and reality of the spectrum}
\author{Emanuela Caliceti}
\address{Dipartimento di Matematica, Universit\`{a} di Bologna, 40126 Bologna, Italy}
\email{emanuela.caliceti@unibo.it}
\author{Sandro Graffi}
\address {Dipartimento di Matematica, Universit\`{a} di Bologna, 40126 Bologna, Italy}
\email{sandro.graffi@unibo.it}

\date{}

\begin{abstract}
{ A class of  non-selfadjoint, $\PT$-symmetric operators is identified  similar to a self-adjoint one, 
thus entailing the reality of the spectrum. 
The similarity transformation is explicitly constructed through the method of the  quantum normal form, whose convergence (uniform with respect to the Planck constant) is proved. Further consequences of the uniform convergence of the quantum normal form are the establishment of an exact quantization formula for the eigenvalues and the integrability of the classical hamiltonian  corresponding to the given $\PT$-symmetric operator.}
\end{abstract} 
\maketitle 
\section{ Introduction and statement of the results}
\setcounter{equation}{0}%
\setcounter{theorem}{0}%
A major mathematical problem in $\PT$-symmetric quantum mechanics (see e.g. \cite{Be3}, \cite{Sp}  \cite{JPhysA}-\cite{Pramana} for recent reviews) is to determine whether or not the spectrum of the $\PT$-symmetric Schr\"odinger operator is real ({\it proper} $\PT$ symmetry \cite{BeBo}).  This is the case, of course, if the given $\PT$-symmetric operator can be conjugated to a self-adoint one through a similarity transformation.  The possibility of such similarity  has been extensively studied (in addition to the relevant references in \cite{Be3}, \cite{Sp}  \cite{JPhysA}-\cite{Pramana}, see also   \cite{Mo1}, \cite{Mo2}, \cite{Mo3} for its examination in an abstract setting). Quite recently, a complete characterization has been obtained of $\PT$-symmetric quadratic Schr\"odinger operators similar to  a self-adjoint one \cite{CGHS}.  

We address in this paper the problem of constructing  such a similarity transformation with the techniques of the Quantum Normal Form ({QNF}) (see e.g. \cite{Sj}, \cite{BGP}),   and  provide a class of $\PT$-symmetric operators for which the procedure works. Namely:  the QNF of the given $\PT$-symmetric Schr\"odinger operator is real and convergent, uniformly  with respect to $\hbar\in [0,1]$.   The convergence of   the QNF  not only provides the similarity with a self-adjoint operator, but has the following  straightforward consequences:
\begin{itemize}
\item[1)]
 It yields an {\it exact} quantization formula for the eigenvalues;
\item[2)]
 Since the  the QNF reduces to the classical normal form (CNF) for $\hbar=0$,  the CNF is convergent as well, and the corresponding classical system is therefore  integrable.
\end{itemize}
Not surprisingly, we are able to prove a result so much stronger than  simple similarity with a self-adjoint operator only for a very restricted class of operators, namely a class of holomorphic, $\PT$-symmetric perturbations of the quantization of the linear diophantine flow over the torus $\T^l$. 

Consider indeed a  {classical} Hamiltonian family, defined in the phase space $\R^l\times \T^l, l=1,2.\dots$, expressed in the action-angle variables $(\xi,x)$, $\xi\in\R^l$, $x\in\T^l$:
\be
\label{Ham}
\H_\ep(\xi,x)=\L_\om(\xi)+\ep \V(\xi,x), \quad \ep\in\R,
\ee
where  $ \L_\om(\xi):=\la\om,\xi\ra$, $\om:=(\om_1,\ldots,\om_l)\in\R^l$, is the Hamiltonian generating the linear quasi-periodic flow  $x_i\mapsto x_i+\om_i t,\;\forall i=1,\dots,l,$ with frequencies $\om_i$ over $\T^l$, and $\V$ is an a priori complex-valued holomorphic function of $(\xi,x)$, assumed to be  $\PT$-symmetric. Namely, if
$\P:\; x\to -x$ denotes the parity operation, i.e. $(\P f)(\xi,x)=f(\xi,-x),\;\forall f\in L^2(\R^l\times \T^l)$ and  ${\mathcal T}: f\to \overline{f}$  the complex conjugation in $L^2(\R^l\times \T^l)$, then 
$$
((\PT)\V)(\xi,x):=\overline{\V}(\xi,-x)={\V}(\xi,x),\quad\forall (\xi,x)\in \R^l\times \T^l.
$$
Writing $\V$ through its uniformly convergent Fourier expansion:
\be
\label{FE}
 \V(\xi,x)=\sum_{q\in\Z^l}\,\V_q(\xi)e^{i\la q,x\ra};\qquad \V_q(\xi)=(2\pi)^{-l/2}\int_{\T^l}\,\V(\xi,x)e^{-i\la q,x\ra}\,dx
 \ee
 the equivalent  formulation of the $\PT$ symmetry in terms of the Fourier coefficients is immediately seen:
 \be
 \label{FCPT}
{\V}_{q}(\xi)=\overline{\V_q(\xi)}, \qquad \forall\,(\xi,q)\in\R^l\times\T^l.
 \ee
 Moreover we assume that
 \be\label{EvenOdd}
 {\V}_{-q}(\xi)=-\V_q(\xi);\qquad \V_q(-\xi)=\V_q(\xi), \qquad \forall\,(\xi,q)\in\R^l\times\T^l,
 \ee
 which ensures that the potential $\V(\xi,x)$ is even in the variable $\xi$ and odd in the variable $x$:
 $$
 \V(-\xi,x) = \V(\xi,x),\qquad \V(\xi,-x) = -\V(\xi,x),\qquad \forall\,(\xi,q)\in\R^l\times\T^l.
 $$
 We denote $V$ the operator in $L^2(\T^l)$ generated by the Weyl quantization of  the symbol $\V$ (see Appendix A.2), namely  the operator acting on $L^2(\T^l)$ in the following way:
 \be
 \label{V}
 (V f)(x):=
\int_{\R^l}\sum_{q\in\Z^l}\widehat{\V}_q(p)e^{i(\la q,x\ra+\la p,q\ra\hbar/2 )}f(x+p\hbar)\,dp, \quad 
 \forall f\in L^2(\T^l),
 \ee
 where
 $$
\widehat{\V}_q(p):=(2\pi)^{-l/2}\int_{\R^l}\,{\V}_q(\xi)e^{-i\la p,\xi\ra }\,d\xi
$$
\vskip 4pt\noindent
is the Fourier transform of  the Fourier coefficient ${\V}_q(\xi)$. 

 Then the quantization of $\H_\ep$ is the $\PT$-symmetric (verification below), non self-adjoint operator in $L^2(\T^l)$ acting as
\be
\label{op}
H(\om,\ep)= i\hbar \la\om,\nabla\ra +\ep V = L(\om,\hbar)+\ep V,\quad L(\om,\hbar):=i\hbar \la\om,\nabla\ra.
\ee
The  \Sc\ operator $H(\om,\ep)$ thus represents a perturbation of the  self-adjoint operator $L(\om,\hbar)$ in  $L^2(\T^l)$,  whose spectrum obviously consists of the eigenvalues  $\lambda_{n,\om}=\hbar\la\om,n\ra$, $n=(n_1,\ldots,n_l)\in\Z^l$, with corrresponding normalized eigenfunctions $\phi_n(x)=(2\pi)^{-l/2}e^{i\la n,x\ra}$.
\begin{remark}
\label{R1}
{\rm By the  assumptions to be specified below $V$ will represent a regular perturbation of $L(\om,\hbar)$. However the spectrum of $L(\om,\hbar)$, although  pure point,  is {\it dense} in $\R$. Therefore the standard (Rayleigh-Schr\"odinger) perturbation theory of quantum mechanics {\it cannot} be applied here because no eigenvalue is isolated, and the approach through the Normal Form is therefore {necessary}, insofar as it represents an alternative method which serves to the purpose.}
\end{remark} 
The statement of the result will profit in clarity by  first sketching the construction of the quantum normal form (QNF) (see e.g. \cite{Sj},\cite{ BGP}, and in this particular context \cite{GP}).  Its purpose  in this connection is to construct a similarity transformation $U(\ep)$ in $L^2(\R^l)$, generated by a continuous operator $W(\ep)$,  $\ds U(\ep)=e^{iW(\ep)/\hbar}$, such that 
\be
\label{sa}
U(\ep)H(\om,\ep)U(\ep)^{-1}=e^{iW(\ep)/\hbar}(L(\om,\hbar)+\ep V)e^{-iW(\ep)/\hbar}=S(\ep)
\ee
where  the similar operator $S(\ep)$ is self-adjoint.  The  procedure goes as follows:
\begin{enumerate}
\item
Look for that particular similarity transformation  $\ds U(\ep)=e^{iW(\ep)/\hbar}$, such that  the transformed operator $S(\ep)$  assumes the form  
\be
\label{serie1}
S(\ep)= L(\om,\hbar) + \sum_{k=1}^{\infty}\ep^k  B_k(\hbar)
\ee
under the additional conditions
\be
\label{diagsa}
 [B_k,L] = 0,\;\qquad B_k=B_k^\ast, \qquad \forall k=1,2,\ldots.
 \ee
 where $ B_k:=B_k(\hbar),\;\forall k$, and $L:=L(\hbar,\om)$. If it can be proved that the series \eqref{serie1} (under the additional conditions \eqref{diagsa}) has a positive convergence radius $\ep^\ast$, then obviously $S(\ep)$ is self-adjoint for $|\ep|<\ep^\ast$, so that its spectrum is real;  moreover, $S(\ep)$ is diagonal on the eigenvector basis of $L(\hbar,\om)$.  The series \eqref{serie1}, assuming the validity of   conditions  \eqref{diagsa}, is called the {\it operator quantum normal form (O-QNF)}. 
  \item
To determine the {O-QNF} we first construct the {QNF} {\it for the symbols} (S-QNF). That is, we first construct for any $k=1,2,\ldots$, the symbol $\B_k(\xi;x;\hbar)$ of the self-adjoint operator $B_k$. The symbol $\B_k$ turns out to be a function only of $\xi$  (depending parametrically on $\hbar$) so that the application of  the Weyl quantization formula (see Appendix A.2) specifies the action of $B_k$:  
$$
 B_k f=\B_k (i\hbar\la \om,\nabla\ra )f=\B_k(L_\om)f, \qquad \forall f\in L^2(\T^l),\quad L_\om:=L=L(\hbar,\om).
 $$
 Hence $[B_k,L_\om]=0,\,\forall k$, and the eigenvalues of $B_k$ are simply $\B_k(n\hbar,\hbar)$, $n\in\Z^l$.  Then the symbol of $S(\ep)$ is
 $$
 \Sigma(\xi,\ep,\hbar)=\L_\om(\xi)+\sum_{k=1}^\infty\B_k(\xi,\hbar)\ep^k
 $$
 provided the series has a non-zero convergence radius.  In that case the eigenvalues of $S(\ep)$, and hence of $H(\om,\ep)$, are clearly given by the following {\it exact} quantization formula:
 \be
 \label{EQF}
 \lambda_n(\ep,\hbar)=\la\om,n\ra\hbar+\sum_{k=1}^\infty\B_k(n\hbar,\hbar)\ep^k,
 \ee
 that is,  by the  symbol $ \Sigma(\xi,\ep,\hbar)$ evaluated at the {\it quantized} values $n\hbar$ of the classical actions $\xi\in\R^l$. 
 Moreover, the spectrum of $S(\ep)$, i.e. of $H(\om,\ep)$, is real if $S(\ep)$ is self-adjoint, namely if $B_k$ is self-adjoint $\,\forall\,k=1,\ldots$; again by the Weyl quantization formula (Appendix A.2), this is true  if  $\B_k(\xi;\hbar)$ is real and bounded  $\,\forall\,k=1, 2,\ldots$. 
 \item By construction, each coefficient $\B_k(\xi,\hbar), k=1,\ldots$, of the S-QNF  turns out to be a smooth function of $\hbar$ near $\hbar=0$, and $\B_k(\xi,0):=\B_k(\xi)$ is just the $k-$term of the classical normal form generated by canonical perturbation theory applied to the classical Hamiltonian $\H_\ep(\xi,x)$. More precisely:
 \be
 \label{can}
 \H_\ep(\xi,x)\sim \L_\om(\xi)+\sum_{k=1}^\infty\,\B_k( \xi)\ep^k
 \ee
where $\sim$ denotes canonical equivalence. Therefore if the convergence  of the S-QNF is {\it uniform} with respect to $\hbar\in [0,1]$ the CNF \eqref{can} is also convergent and therefore the classical hamiltonian $\H_\ep(\xi,x)$ is integrable because the equivalent hamiltonian depends only on the actions.
\end{enumerate}
We can now proceed to the precise statement of the results.  First we describe the assumptions.  Consider again the operator 
\begin{eqnarray*}
&&
 L(\om,\hbar)\psi = i\hbar \la\om,\nabla\ra\psi
=-i\hbar\left[\om_1\frac{\partial}{\partial x_1}+\ldots+\om_l\frac{\partial}{\partial x_l}\right]\psi , \quad
\forall\psi\in  D(L_\om)=H^1(\T^l);
\\
&&
H^1(\T^l):=\{\psi=\sum_{n\in\Z^l}\,\psi_ne^{i\la n,x\ra}\in L^2(\T^l)\,:\,\sum_{n\in\Z^l}\,|n|^2\,|\psi_n|^2 <+\infty\}
\end{eqnarray*}
The first assumption is :
\par\noindent
(A1) {\it The frequencies $\om=(\om_1,\ldots,\om_l)$ are} diophantine, {\it i.e. $\exists\gamma>0,\; \tau>l$ such that:}
\be
\label{DC}
|\la\om,q\ra|^{-1}\leq \gamma |q|^{\tau}, \quad q \in\Z^l, \; q\neq 0.
\ee
Remark that \eqref{DC} entails that all the eigenvalues $\lambda_{n,\om}=\la n,\om\ra\hbar$ of $L(\om,\hbar)$ are simple.

Let now $(t,x)\mapsto \V(t,x)$ be a complex-valued smooth function defined on $\R\times\T^l$, i.e. $\V\in C^{\infty}(\R\times\T^l;\C)$. Write its Fourier expansion:
\be
\label{FV}
\V(t,x)=\sum_{q\in\Z^l}\,\V_{q}(t)e^{i\la q,x\ra}, \quad \V_{q}(t):=(2\pi)^{-l/2}\int_{\T^l}\V(t,x)e^{-i\la q,x\ra}\,dx
\ee
and define the functions $\V_\om(\xi,x):\R^l\times\T^l\to\C$ in the following way:
\be
\label{Vom}
\V_\om(\xi,x):=\V(\la\omega,\xi\ra,x)=\sum_{q\in\Z^l}\,\V_{\om,q}(\xi)e^{i\la q,x\ra}, \qquad \V_{\om,q}(\xi):=\V_q(\la\om,\xi\ra).
\ee
Now consider the
space Fourier transform of $\V_{q}(t), q\in\Z^l$:
$$
\widehat{\V}_q(p):=\frac{1}{\sqrt{2\pi}}\ds\int_{\R}\,\V_q(t)e^{-ipt}\,dt,\quad p\in\R.$$

Then (see formula (\ref{(A1)})) the Weyl quantization of $\V_\om(\xi,x)$ is the operator in $L^2(\T^l)$ acting as follows:
$$
(V_\om f)(x)=\int_{\R}\sum_{q\in\Z^l}\widehat{\V}_q(p)
e^{i(\la q,x\ra+\hbar p\la \om,q\ra/2)}f(x+\hbar p\om)\,dp, \quad f\in L^2(\T^l). 
$$
$V_\om$ is actually a continuous operator in $L^2(\T^l)$ (see Appendix, Remark \ref{Ra1}(d)) by virtue of  our second assumption, namely:
\vskip 5pt\noindent
(A2) {\it Let the diophantine constants $\gamma$ and $\tau$ be such that }
$$
\gamma\tau^\tau(\tau+2)^{4(\tau+2)}<\frac12
$$
{\it and let there exist $\rho>2$ such that }
\be
\label{normarho} 
  \|\V_\om\|_{\rho}:=\sum_{q\in\Z^l}\,e^{\rho |q|} \int_{\R}e^{\rho |p|}|\widehat{\V}_q(p)|\,dp<+\infty.
 \ee
 \vskip 4pt\noindent
\begin {remark}\label{R2}
{}
\par\noindent
\begin{itemize}
{\rm \item[(i)]
 Actually, by formula (A.6), $\|V_\om\|_{L^2\to L^2}\leq \|\V_\om\|_{\rho}$.  Moreover, assumption (A2) makes $\V_\om$ a holomorphic function of $(\xi,x)$ in $\C_{\rho}^{2l}:=\{(\xi,x)\in\C^{2l}\,:\,|\im\xi_i |<\rho; \;|\im x_i |<\rho,\;\forall i=1,\dots,l\} $. 
 \item[(ii)]
 As discussed in \cite{GP}, $\V(t,x)$ must depend explicity on $t$ if $l>1$ to make the problem a nontrivial one. Once more by (A2), formula (\ref{normarho}), $\V(t,x)$ vanishes exponentially fast as $|t|\to\infty$ uniformly w.r.t. $x\in\T^l$.}
 \end{itemize}
 \end{remark}
Our third assumption concerns the $\PT$-symmetry, and is formulated as follows (see (\ref {FCPT}) and (\ref{EvenOdd})):
\vskip 5pt\noindent
(A3) {\it The Fourier coefficients $\V_{\om,q}(\xi)$ enjoy the following symmetry properties:}
\be
\label{PTSP}
\V_{\om,q}(\xi)=\overline{\V_{\om,q}(\xi)};\quad\V_{\om,-q}(\xi)=-\V_{\om,q}(\xi);\quad \V_{\om,q}(-\xi)=\V_{\om,q}(\xi),\quad \forall (\xi,q)\in\R^l\times\T^l.
\ee
\begin{remark}\label{R3}
{\rm Clearly (A3) entails $\V_\om(\xi,-x)=-\V_\om(\xi,x)$ and 
$$
((\PT)\V_\om)(\xi,x)=(\PT)\ds(\sum_{q\in\Z^l}\V_{\om,q}(\xi)e^{i\la q,x\ra}\ds)=\V_\om(\xi,x),\qquad \forall(\xi,x)\in\R^l\times\T^l,
$$
that is, $\V_\om(\xi,x)$ is a $\PT$-invariant function, odd with respect to $x$. Moreover from (\ref{PTSP}) one can easily obtain
$\widehat{\V}_{\om,q}(-p)=\widehat{\V}_{\om,q}(p)\in\R,\;\forall p\in\R^l,\,\forall q\in\Z^l$.
This entails that $V:=V_\om$ is a $\PT$-symmetric operator in $L^2(\T^l)$, i.e. $[V,\PT]=0$. We have indeed}
\begin{eqnarray*}
(\PT)(Vf)(x)
&=& \int_{\R}\sum_{q\in\Z^l}\widehat{\V}_{\om,q}(p)
e^{i\la q,x\ra-i\hbar p\la \om,q\ra/2}\overline{f}(-x+\hbar p\om)\,dp
\\
&
=& \int_{\R}\sum_{q\in\Z^l}\widehat{\V}_{\om,q}(p)
e^{i(\la q,x\ra+\hbar p\la \om,q\ra/2)}\overline{f}(-x-\hbar p\om)\,dp
\\
&=&\ds \int_{\R}\sum_{q\in\Z^l}\widehat{\V}_{\om,q}(p)
e^{i(\la q,x\ra+\hbar p\la \om,q\ra/2)}(\PT f)(x+\hbar p\om)\,dp
\\
&=&V(\PT f)(x)\,,\qquad  \forall f\in L^2(\T^l),\;\forall x\in\T^l.
\end{eqnarray*}
\end{remark}
To sum up, the operator family acting as
$$
H(\ep) =  i\hbar \la\om,\nabla\ra + \ep V
$$ 
and defined on 
$D(H(\ep))=H^1(\T^l)$ has pure-point spectrum denoted $
\sigma(H(\ep))$, and we will prove that it consists of a sequence of non-isolated eigenvalues denoted
$\{\l_n(\hbar,\ep):\; n\in\Z^l\}$.
The symbol of $H(\ep)$ is the Hamiltonian family defined on $\R^l\times\T^l$:
$$
\H_\ep(\xi,x)=\la\om,\xi\ra + \ep \V_\om(\xi,x) = \L_\om(\xi)+\ep \V_\om(\xi,x).
$$
\vskip 5pt
We can now state the main result of the paper.
\begin{theorem}
\label{mainth}
 Under Assumptions (A1-A3),  there exists $\ep_0>0$ independent of $\hbar\in [0,1]$ 
such that for $|\ep|<\ep_0$ the spectrum of $H(\ep)$ is given by the exact quantization formula:
\bea
\label{EQF1}
&&
\lambda_n(\hbar,\ep)=\la\om,n\ra\hbar+\B(n\hbar,\hbar;\ep), \quad n\in\Z^l
\\
&&
\label{serieB}
 \B(n\hbar,\hbar;\ep):=\ds\sum_{k=1}^\infty\,\B_k(n\hbar,\hbar)\ep^k
 \eea
where
\begin{enumerate}
\item  $\B_k(\xi,\hbar)\in C^\infty(\R^l\times [0,1])$ is real-valued, $k=1,2,\ldots\;$;
\item $\B_{2s+1}=0$, $s=0,1,\ldots\;$;
\item The series \eqref{serieB} converges uniformly with respect to $(\xi,\hbar)\in\R^l\times [0,1]$;
\item $\B_k(n\hbar,\hbar)$ is obtained from the Weyl quantization formula applied to $\B_k(\xi,\hbar)$, which is the symbol of the  operator $B_k$, the term of order $k$ of the QNF.
\end{enumerate}
\end{theorem}
\begin{corollary}\label{C1}
Let $|\ep|<\ep_0$. Then the operator $H(\om,\ep)$ is  similar to the selfadjoint operator 
$$
S(\ep):=L(\om,\hbar)+\sum_{k=1}^\infty\,B_k(\hbar).
$$
\end{corollary}
\begin{remark}\label{R4}
{\rm The explicit construction of the bounded operator $W(\ep)$ realizing the similarity $U=U(\om, \ep, \hbar)= e^{iW(\ep)/\hbar}$ is described in the proof of Theorem \ref{mainth}.}
\end{remark}
A straightforward consequence of the uniformity (with respect to $\hbar\in [0,1]$) of the convergence of the QNF is a convergence result for the  corresponding CNF, valid for a class of
 $\PT$-symmetric,
non-holomorphic perturbations of 
non-resonant harmonic oscillators. 
Consider indeed the {inverse transformation}  into action-angle variables
$$
{\mathcal C}(\xi,x)=(\eta,y):=  \left\{\begin{array}{c} \eta_i=-\sqrt{\xi_i}\sin x_i 
\\
\\
 y_i=\sqrt{\xi_i}\cos x_i \end{array}\right.\quad i=1,\ldots,l
$$
It is defined only on $\R_+^l\times \T^l$ and does not preserve the regularity at the origin.
On the other hand,  ${\mathcal C}$ is an {analytic}, {canonical} map between $\R_+^l\times\T^l$ and $\R^{2l}\setminus\{0,0\}$.
\nl
Then
$$
(\H_\ep \circ {\mathcal C}^{-1})(\eta,y)= \sum_{s=1}^l\om_s(\eta^2_s+y_s^2)+\ep (\V\circ {\mathcal C}^{-1})(\eta,y)
$$
$$
:=\P_0(\eta,y)+\ep \P_1(\eta,y)
$$
where for $\;(\eta,y)\in\R^{2l}\setminus\{0,0\}$
$$
\P_1(\eta,y)=(\V\circ {\mathcal C}^{-1})(\eta,y)=\P_{1,R}(\eta,y)+\P_{1,I}(\eta,y),  
$$
$$
\P_{1,R}(\eta,y)=\frac12\sum_{k\in\Z^l}(\Re{\V}_k\circ {\mathcal C}^{-1})(\eta,y)\prod_{s=1}^l \left(\frac{\eta_s-iy_s}{\sqrt{\eta^2_s+y_s^2}}\right)^{k_s}
$$
$$
\P_{1,I}(\eta,y)=\frac12\sum_{k\in\Z^l} (\Im{\V}_k\circ {\mathcal C}^{-1})(\eta,y)\prod_{s=1}^l \left(\frac{\eta_s-iy_s}{\sqrt{\eta^2_s+y_s^2}}\right)^{k_s}
$$
\begin{corollary}\label{C2}
The Birkhoff normal form of $\H_\ep$ is {real} and uniformly convergent on any compact of  $\R^{2l}\setminus\{0,0\}$ if $|\ep|<\ep_0$. Hence the system is integrable.
\end{corollary}
\section{Proof of the results}
\setcounter{equation}{0}%
\setcounter{theorem}{0}%
{\it Proof of Theorem \ref{mainth}.} Under the present conditions,  statements (3) and (4) are proved in \cite{GP}, as well as the smoothness  of $\B_k(\xi,\hbar)$ asserted in (1). The assertions left to prove are therefore the reality statement (1), $B_k(\xi,\hbar)=\overline{B}_k(\xi,\hbar)$, $\forall\,(\xi,\hbar)\in\R^l\times [0,1]$, and the even nature  of the QNF (2), $\B_{2s+1}=B_{2s+1}=0,\,\forall s=0, 1, \dots$.  This requires a detailed examination of the structure of the QNF, whose construction we now recall in Subsection 2.1. In Subsection 2.2 we describe the inductive argument proving the reality assertion, and the symmetry argument proving  he vanishing of the odd terms. 
\subsection{The Quantum Normal Form: 
the formal construction} 
(We follow Sj\"ostrand \cite{Sj} and Bambusi-Graffi-Paul \cite{BGP}).

Given $H(\ep) =  L(\om,\hbar) + \ep V$ in $L^2(\T^l)$, look for a similarity transformation $U=U(\om,\ep,\hbar)$, in general {non unitary} ($W(\ep)\neq W(\ep)^\ast$):
$$
U(\om,\ep,\hbar)=e^{i W(\ep)/\hbar}:
L^2(\T^l)\leftrightarrow L^2(\T^l)
$$
such that
\be
\label{2}
S(\ep):=UH(\ep) U^{-1}=L(\om,\hbar)+\ep B_1+\ep^2 B_2+\ldots 
= L(\om,\hbar) +  \sum_{k=1}^{\infty}B_k\ep^k
\ee
under the requirement: 
$$
[B_k,L]=0, \qquad \forall k.
$$
Recall the formal commutator expansion
\be\label{SH}
S(\ep)=e^{i W(\ep)/\hbar}H(\ep) e^{-i W(\ep)/\hbar}=\sum_{k=0}^\infty H_k
\ee
$$
H_0:=H(\ep),\quad
H_k:=\frac{[W(\ep),H_{k-1}]}{i\hbar k}, \qquad k\geq 1
$$
and look for $W(\ep)$ in the form of a power series expansion in $\ep$: \quad
$W(\ep)=\ep W_1+\ep^2W_2+\ldots.$
\nl
Then \eqref{2} becomes:
\be\label{SB}
S(\ep)=\sum_{k=0}^{\infty}\ep^k  B_k
\ee
where
\be\label{BW}
B_0=L(\om,\hbar);\quad {B}_k:=\frac{[W_k,L]}{i\hbar}+V_k\,,\qquad k\geq 1,
\ee
 $V_1\equiv V$ and
 \begin{eqnarray}\label{VW}
V_k &=&\sum_{r=2}^k\frac{1}{r!}\sum_{{j_1+\ldots+j_r=k}\atop {j_s\geq
1}}\frac{[W_{j_1},[W_{j_2},\ldots,[W_{j_r},L]\ldots]}{(i\hbar)^r}\nonumber\\
\\
&+&
\sum_{r=1}^{k-1}\frac{1}{r!}\sum_{{j_1+\ldots+j_r=k-1}\atop {j_s\geq
1}}\frac{[W_{j_1},[W_{j_2},\ldots,[W_{j_r},V]\ldots]}{(i\hbar)^r}. \nonumber
\end{eqnarray}
$V_k$ depends on $W_1,\dots, W_{k-1}$, but not on $W_k$. Thus we get the recursive homological equations:
\be
\label{3}
\frac{[W_k,L]}{i\hbar} +V_k=B_k, \qquad [L,B_k]=0.
\ee
To solve \eqref{3} for the two unkowns $ B_k, W_k$, we look for their symbols and then apply the Weyl quantization formula. First recall (see e.g. \cite{Fo} or \cite{Ro}) that the symbol of the commutator $[F,G]/i\hbar$ of two operators $F$ and $G$ is the {\it Moyal bracket} $\{\F,\G\}_M$ of the symbols $\F=\F(\xi,x,\hbar)$ of $F$ and $\G=\G(\xi,x,\hbar)$ of $G$, where $\{\F,\G\}_M$ is defined through its Fourier representation
\vskip 6pt\noindent
\be\label{M1}
\{\F,\G\}_M(\xi,x;\hbar) = \int_{\R^l}\sum_{q\in\Z^l}\widehat{(\{\F,\G\}_M)}_q(p,\hbar)
e^{i(\la p,\xi\ra\la + \la q,x\ra)}\,dp
\ee
\vskip 4pt\noindent
and
\be\label{M2}
\widehat{(\{\F,\G\}_M)}_q(p,\hbar) = \frac{2}{\hbar} \int_{\R^l}\sum_{q'\in\Z^l}\widehat{\F}_{q-q'}(p-p',\hbar)\widehat{\G}_{q'}(p',\hbar)\sin[ \frac{2}{\hbar}(\la p',q\ra-\la p,q'\ra)]\,dp'\,.
\ee
Notice that  $\{\F,\G\}_M = -\{\G,\F\}_M$.
The above equations (\ref {SH})-(\ref {VW}) become, once written for the symbols:
\be\label{SH1}
 \Sigma(\ep)=\sum_{k=0}^\infty {\H}_k
\ee
$$
 {\H}_0:=\L_\om+\ep\V,\quad  {\H}_k:=\frac{\{\W(\ep),{\H}_{k-1}\}_M}{ k}, \;k\geq 1,
 $$
where 
$\W(\ep)=\ep \W_1+\ep^2\W_2+\ldots$,
\be\label{SB1}
 \Sigma(\ep)=\ds\sum_{k=0}^{\infty}\ep^k  \B_k
\ee
 and
\be\label{BW1}
\B_0=\L_\om=\la\om,\xi\ra;\quad \B_k =\{\W_k,{\L_\om} \}_M+\V_k,\; k\geq1,\quad
\V_1\equiv \V
\ee
\begin{eqnarray}\label{VW1}
\V_k &=& \sum_{r=2}^k\frac{1}{r!}\sum_{{j_1+\ldots+j_r=k}\atop {j_s\geq
1}}\{\W_{j_1},\{\W_{j_2},\ldots,\{\W_{j_r},\L_\om\}_M\ldots\}_M\}_M 
\\
\nonumber
&+&\sum_{r=1}^{k-1}\frac{1}{r!}\sum_{{j_1+\ldots+j_r=k-1}\atop {j_s\geq
1}}\{\W_{j_1},\{\W_{j_2},\ldots,\{\W_{j_r},\V\}_M\ldots\}_M\}_M , \quad k>1
\end{eqnarray}
Therefore the symbols $\W_k$ and $\B_k$ of $W_k$ and $B_k$ can be recursively found solving the homological equation:
\be
\label{5}
\{\W_k,\L_\om\}_M +\V_k=\B_k, \qquad k=1,\ldots
\ee
under the condition:
\be
 \label{6}
 \{\L_\om,\B_k\}_M =0.
 \ee
Here
$$
\W_k=\W_k(\xi,x;\hbar),\;\V_k=\V_k(\xi,x;\hbar),\;\B_k=\B_k(\xi,x;\hbar).
$$
Notice that, in view of Theorem \ref{Ta1} in Appendix, \eqref{6} is immediately satisfied if $\B_k=\B_k(\xi;\hbar)$ does not depend on $x$. Moreover, by Theorem \ref{Ta1}(2), since
$\L_\om=\L_\om(\xi)=\la\om,\xi\ra$ is linear in $\xi$, we have
$$
\{\W_k,\L_\om\}_M = \{\W_k,\L_\om\} = -\la\nabla_x\W_k,\om\ra
$$
and \eqref{5} becomes 
\be
\label{7}
-\la\nabla_x\W_k(\xi,x),\om\ra + \V_k(\xi,x;\hbar) = \B_k(\xi;\hbar). 
\ee
Write  now $W_k(\xi,x;\hbar)$ and $\V_k(\xi,x;\hbar)$ under their Fourier series representation, respectively: 
 $$
 \W_k(\xi,x;\hbar)=\sum_{q\in\Z^l}\W_{k,q}(\xi;\hbar)e^{i\la q,x\ra}, \qquad \V_k(\xi,x;\hbar)=\sum_{q\in\Z^l}\V_{k,q}(\xi;\hbar)e^{i\la q,x\ra}.
$$ 
Then \eqref{7} in turn becomes:
\be\label{8}
-i\sum_{q\neq 0}\la q,\om\ra\W_{k,q}(\xi;\hbar)e^{i\la q,x\ra} + \sum_{q\in\Z^l}\V_{k,q}(\xi;\hbar)e^{i\la q,x\ra} = \B_k(\xi;\hbar)
\ee
whence, imposing the equality of the Fourier coefficients of both sides, we obtain the solutions
\be
\label{sol}
\B_k(\xi,\hbar) = \V_{k,0}(\xi,\hbar), \qquad
\W_{k,q}(\xi,\hbar) = \frac{\V_{k,q}(\xi,\hbar)}{i\la q,\om\ra},\quad \forall q\neq 0.
\ee
\subsection{Reality of $\B_k$: the inductive argument} 
Denote now $\V_1\equiv \V = \V_\om$.
Since $\V_{\om,q}(\xi)$ is real $\forall q\in\Z^l$ by assumption, we have
$$
\B_1(\xi,\hbar) = \V_{\om,0}(\xi)\in\R
 $$ 
 and
\be\label{solution}
\W_{1,q}(\xi,\hbar) = \frac{\V_{\om,q}(\xi)}{i\la q,\om\ra}\in i\R, \quad\forall q\neq 0.
\ee
Moreover, since no requirement is asked on $\W_{1,0}$, we can choose $\W_{1,0}=0$. 
Now assume inductively:
\newline
(${\bf A_1}$)
$
 \V_{j,q}(\xi,\hbar)\in\R,\quad\forall j=1,\dots,k-1,\; \forall q\in\Z^l;
 $
 \newline
${\bf (A_2)}$ we can choose $\W_{j,0}=0,\;\forall j=1,\dots,k-1.$ 
\nl
Remark that (${\bf A_1}$) entails
\be
\label{ReIm}
 \W_{j,q}(\xi,\hbar)) = \frac{\V_{j,q}(\xi,\hbar)}{i\la q,\om\ra}\in i\R\,,\;\qquad  \B_j(\xi,\hbar) = \V_{j,0}\in\R, \quad \forall j=1,\dots,k-1.
\ee
Then the following assertions hold:
\nl
$ {\bf (R_1)}$
 $
  \V_{k,q}(\xi,\hbar)\in\R,\; \forall q\in\Z^l;
  $
  \nl
$ {\bf (R_2)}$ we can choose $\W_{k,0}=0$.
\nl
Remark that $ {\bf (R_1)}$ entails
\be
\label{ReIm1}
 \W_{k,q}(\xi,\hbar) = \frac{\V_{k,q}(\xi,\hbar)}{i\la q,\om\ra}\in i\R\,; \qquad \B_k(\xi) = \V_{k,0}\in\R.
 \ee
In order to prove $ {\bf (R_1)}$ consider the Fourier expansion of $\V_k$ given by (\ref{VW1}) 
\begin{eqnarray*}
&&
\V_k = \sum_{r=2}^k\frac{1}{r!}\sum_{{j_1+\ldots+j_r=k}\atop {j_s\geq
1}}\{\W_{j_1},\{\W_{j_2},\ldots,\{\W_{j_r},\L_\om\}_M\ldots\}_M 
\\
&&
+\sum_{r=1}^{k-1}\frac{1}{r!}\sum_{{j_1+\ldots+j_r=k-1}\atop {j_s\geq
1}}\{\W_{j_1},\{\W_{j_2},\ldots,\{\W_{j_r},\V\}_M\ldots\}_M
\\
&&
=\sum_{q\in\Z^l}\,\V_{k,q}(\xi,\hbar)e^{i\la q,x\ra}. 
\end{eqnarray*}
\vskip 5pt\noindent
By \eqref{ReIm},  the Fourier coefficients $\W_{j_s,q}$ of each term $\W_{j_s},\; s=1,\dots,r,$ are purely imaginary, and by Theorem \ref{Ta1}(3)  each Moyal bracket generates another factor $i$. 
Therefore 
$$
\Big(\sum_{{j_1+\ldots+j_r=k}\atop {j_s\geq
1}}\{\W_{j_1},\{\W_{j_2},\ldots,\{\W_{j_r},\L_\om\}_M\ldots\}_M\Big)_q(\xi,\hbar) =(i)^{2r}a_{k,q}(\xi,\hbar), \quad  a_{k,q}(\xi,\hbar)\in\R
$$
$$
\Big(\sum_{{j_1+\ldots+j_r=k-1}\atop {j_s\geq
1}}\{\W_{j_1},\{\W_{j_2},\ldots,\{\W_{j_r},\V\}_M\ldots\}_M\Big)_q(\xi,\hbar)=(i)^{2r}b_{k,q}(\xi,\hbar), \quad  b_{k,q}(\xi,\hbar)\in\R
$$
\vskip 4pt\noindent
and, as a consequence:
$$
\V_{k,q}(\xi, \hbar)= (i)^{2r}[a_{k,q}(\xi,\hbar)+a_{k,q}(\xi,\hbar)]=(-1)^r[a_{k,q}(\xi,\hbar)+a_{k,q}(\xi,\hbar)]\in\R, \quad \forall q\in\Z^l.
$$
Hence $\B_k(\xi,\hbar)=\V_{k,0}\in\R$. Moreover, the homological equation (\ref {8}) does not involve $\W_{k,0}$, therefore we can always take $\W_{k,0}=0$. This concludes the proof of the induction, and thus of  Assertion (1) of Theorem \ref{mainth}.
\subsection{Vanishing of the odd terms $\B_{2s+1}$}
Let us now prove Assertion (2) of Theorem \ref{mainth}.  This will yield
$$
\Sigma(\ep)=\B(\xi;\hbar)=\L_\om(\xi)+\ep^2\B_2(\xi,\hbar)+\ep^4\B_4(\xi,\hbar)+\dots. 
$$
{To see this, first recall that $\V_\om(\xi,x)$ is odd in $x$:
$\V_\om(\xi,-x)=-\V_\om(\xi,x)$, and let $\M$ denote the set of  functions $f:\T^l\to\C$ with a definite parity (either even or odd). Moreover,  $\forall\, f\in\M$ define 
$$
Jf=  \left\{\begin{array}{c} +1, \quad {\rm if} f\;{\rm is\; even},
\\
\\
-1, \quad {\rm if} f\;{\rm is \;odd}. \end{array}\right.
$$
Then $Jf=1$ if and only if $f_q=f_{-q}$ and $Jf=-1$ if and only if $f_q=-f_{-q}, \forall  q\in\Z^l$. By assumption $\V_{\om,q}(\xi)=-\V_{\om,-q}(\xi), \forall  q\in\Z^l, \forall  \xi\in\R^l$, i.e. $J\V_{\om}(\xi)=1$, and by  (\ref{solution})
$$
J\W_1(\xi,\hbar) = 1,\qquad \forall(\xi,\hbar)\in\R^l\times[0,1].
$$
Now we can prove by induction that
\be\label{J}
J\V_k=(-1)^k,\qquad \forall k=1,2,\dots
\ee
whence $J\V_{2s+1}=1$, i.e. $\V_{2s+1}(\xi,x,\hbar)$ is odd in $x$, which entails $\B_{2s+1}=\V_{2s+1,0}=0, \forall s=0,1,\dots$. To prove (\ref{J}) inductively first notice that $J\V_{1}=J\V_{\om}=1$ and then let us assume that 
$$
J\V_{j}=(-1)^j,\qquad \forall j=1,\dots,k-1.
$$
Then by (\ref{sol})
$$
J\W_{j}=(-1)^{j+1},\qquad \forall j=1,\dots,k-1.
$$
Let us examine the parity of the first summand in the r.h.s. of (\ref{VW1}), making use of Theorem \ref{Ta1}(4):
$$
J(\{\W_{j_1},\{\W_{j_2},\dots\{\W_{j_r},\L_{\om}\}_M\dots\}_M\}_M) = (-1)^r(-1)^{j_1+1}\dots(-1)^{j_r+1}=(-1)^k
$$
since $J\L_{\om}=1$ and $j_1+\dots+j_r=k$.
Similarly for the second summand in the r.h.s. of (\ref{VW1}) we have
$$
J(\{\W_{j_1},\{\W_{j_2},\dots\{\W_{j_r},\V\}_M\dots\}_M\}_M) = (-1)^{r+1}(-1)^{j_1+1}\dots(-1)^{j_r+1}=(-1)^k
$$
since $J\V=-1$ and $j_1+\dots+j_r=k-1$. This completes the proof of Assertion (2) and hence of Theorem \ref{mainth}.
\newline
{\it Proof of Corollary \ref{C1}.}
It is proved in \cite{GP} that the convergence of the S-QNF
$$
\Sigma(\ep)=\L_\om(\xi)+\sum_{k=1}^\infty \B_k(\xi,\hbar)\ep^k
$$
takes place in the $\|\cdot\|_{\rho/2}$-norm, where $\|\cdot\|_\rho$ is the norm defined in \eqref{normarho}.    Since (Remark \ref{A2}(b) and Appendix A.2) the  $\|\cdot\|_{\rho/2}$-norm majorizes the operator norm in $L^2(\T^l)$ of the corresponding Weyl-quantized operators, we can conclude that 
$$
S(\ep)=L(\om,\hbar)+\sum_{k=1}^\infty\,B_k\ep^k,\qquad B_{2s+1}=0, \quad\forall s=0,1,\dots,
$$
where the convergence takes place in the operator norm sense. Since $B_k=B_k^\ast$, $S(\ep)=S(\ep)^\ast$ and the similarity between $H_\ep$ and a self-adjoint operator is therefore proved.
\subsection{Proof of Corollary \ref{C2}}
By the uniform convergence of the S-QNF with resepct to $\hbar\in [0,1]$, it is enough 
to check that $\B_k(\xi,0)$ is the $k-$th coefficient of the CNF for $\H_\ep(\xi,x)$. 
\newline
Under the present regularity assumptions it is known (see e.g.\cite{Sj}, \cite{BGP}) that, for each $k$,   $\W_k(\xi,x;\hbar),\; $ $\B_k(\xi;\hbar),\;$ $ \V_k(\xi,x;\hbar)$ admit an asymptotic expansion in powers of $\hbar$ near $\hbar=0$:
$$ 
\W_k(\xi;x;\hbar)\sim \ds\sum_{j=0}^\infty\,\W_k^{(j)}(\xi,x)\hbar^j;\quad 
\B_k(\xi;\hbar)\sim \ds\sum_{j=0}^\infty\,\B_k^{(j)}(\xi)\hbar^j\quad 
\V_k(\xi;x;\hbar)\sim \ds\sum_{j=0}^\infty\,\V_k^{(j)}(\xi,x)\hbar^j.
$$
Let us now prove that the terms of order zero in the above expansions, namely the {\it principal symbols} of $\W_k(\xi,x;\hbar),\; $ $\B_k(\xi;\hbar),\;$ $ \V_k(\xi,x;\hbar)$, respectively 
$$
w_k:=\W_k^{(0)},\quad b_k=\B_k^{(0)},\quad v_k=\V_k^{(0)}
$$ 
coincide with the coefficients of order $k$ of the CNF generated by the Hamiltonian family
$\H_\ep(\xi,x)=\L_\om(\xi)+\ep \V_\om(\xi,x)$. In fact, the recursive homological equations (\ref{5}) and (\ref{6})
$$
\{\W_k,\L\}_M +\V_k=\B_k, \qquad  \{\L,\B_k\}_M =0,\quad k=1,\ldots
$$
evaluated at $\hbar=0$ become
$$
\{w_k,\L\} + v_k=b_k, \qquad \{\L,b_k\}=0, \qquad 
\v_1\equiv v\equiv \V
$$
\bea
\label{9}
v_k &=& \ds\sum_{r=2}^k\frac{1}{r!}\ds\sum_{{j_1+\ldots+j_r=k}\atop {j_s\geq
1}}\{w_{j_1},\{w_{j_2},\ldots,\{w_{j_r},\L\}\ldots\}
\\
\nonumber
&+&
\ds\sum_{r=1}^{k-1}\frac{1}{r!}\ds\sum_{{j_1+\ldots+j_r=k-1}\atop {j_s\geq
1}}\{w_{j_1},\{w_{j_2},\ldots,\{w_{j_r},v\}\ldots\} 
\eea
where $\{f,g\}$ denotes the Poisson bracket of two observables $f, g\in C^{\infty}(\R^l\times\T^l)$.
Let us check that this is exactly the recurrence defined by 
{canonical perturbation theory} generated by the Lie transformation algorithm.  
Look indeed for an $\ep$-dependent family of smooth canonical maps 
$\Phi_\ep: \R^l\times \T^l \leftrightarrow \R^l\times \T^l$,
$ (\xi,x)\mapsto (\eta,y)=\Phi_\ep(\xi,x)$
such that
\be
\label{10}
\H_\ep\circ \Phi_\ep^{-1}(\xi,x) =\L(\xi)+\ep b_1(\xi)+\ep^2 b_2(\xi)+\ldots
\ee
Look for $\Phi_\ep$ as the time 1 flow of a smooth Hamiltonian family $w_\ep(\xi,x)$, the  
{\ generating function}. Then
\be
\label{11}
\H_\ep\circ \Phi_\ep^{-1}(\xi,x)
=\H_\ep(\xi,x)+\ds\sum_{s=1}^\infty\,\{w_\ep^{(1)},\{w_\ep^{(2)},\ldots\{w_\ep^{(s)},\L\}\ldots\}\}
\ee
where $w_\ep^{(r)}=w_\ep,\;\forall r=1,2,\dots$. If we set
$$
w_\ep=\ep w_1+\ep^2 w_2+\ldots
$$
 and require equality between \eqref{10} and \eqref{11} we obtain
$$
{b}_k=\{w_k,\L\}+v_k,\quad k\geq 1, \;v_1\equiv v \equiv \V
$$
\begin{eqnarray*}
v_k &= &\sum_{r=2}^k\frac{1}{r!}\sum_{{j_1+\ldots+j_r=k}\atop {j_s\geq
1}}\{w_{j_1},\{w_{j_2},\ldots,\{w_{j_r},\L\}\ldots\}
\\
&+&
\sum_{r=1}^{k-1}\frac{1}{r!}\sum_{{j_1+\ldots+j_r=k-1}\atop {j_s\geq
1}}\{w_{j_1},\{w_{j_2},\ldots,\{w_{j_r},v\}\ldots\}
\end{eqnarray*}
Condition $\{\L,b_k\}=0$ follows from the fact that both $\L(\xi)$ and $b_k(\xi)$ do not depend on $x$. This concludes the proof of the corollary.
\begin{appendix}
\section{Moyal brackets and the Weyl quantization }
\setcounter{equation}{0}%
\setcounter{theorem}{0}%
\subsection{Moyal brackets}
\begin{theorem}\label{Ta1}
Let $\F=\F(\xi,x;\hbar)$ and $\G=\G(\xi,x;\hbar)$ belong to $C^{\infty}(\R^l\times\T^l\times[0,1]; \C)$ and vanish exponentially fast as $|\xi|\to\infty$, uniformly with respect to $(x,\hbar)\in\T^l\times[0, 1]$. Consider their Fourier representation
\begin{eqnarray*}
\F(\xi,x;\hbar)=\int_{\R^l}\sum_{q\in\Z^l}\widehat{\F}_q(p;\hbar)e^{i(\la p,\xi\ra +\la q,x\ra)}\,dp
\\
\G(\xi,x;\hbar)=\int_{\R^l}\sum_{q\in\Z^l}\widehat{\G}_q(p;\hbar)e^{i(\la p,\xi\ra +\la q,x\ra)}\,dp\,,
\end{eqnarray*}
where
\begin{eqnarray*}
\F_q(\xi,\hbar) = (2\pi)^{-l/2}\int_{\T^l}\F(\xi,x,\hbar)e^{-i\la q,x\ra}\,dx
\\
\G_q(\xi,\hbar) = (2\pi)^{-l/2}\int_{\T^l}\G(\xi,x,\hbar)e^{-i\la q,x\ra}\,dx
\end{eqnarray*}
and
\begin{eqnarray*}
\widehat{\F}_q(p;\hbar) = (2\pi)^{-l/2}\int_{\R^l}\F_q(\xi,\hbar)e^{-i\la p,\xi\ra}\,d\xi
\\
\widehat{\G}_q(p;\hbar) = (2\pi)^{-l/2}\int_{\R^l}\G_q(\xi,\hbar)e^{-i\la p,\xi\ra}\,d\xi\,.
\end{eqnarray*}
Then the following assertions hold:
\begin{enumerate}
\item[(1)]
If both $\F$ and $\G$ do not depend on $x$, i.e. $\F(\xi,x;\hbar)=\F(\xi;\hbar)$ and $\G(\xi,x;\hbar)=\G(\xi;\hbar) $, then $\{\F,\G\}_M \equiv 0$.
\item[(2)]
If $\G(\xi,x;\hbar)=\la\om,\xi\ra$, for a given constant vector $\om\in\R^l$, i.e. $\G$ does not depend on $x$ and is linear in $\xi$, then 
$$
\{\F,\G\}_M = \{\F,\G\} = -\la\nabla_x\F,\om\ra\,.
$$
\item[(3)]
Consider the Fourier expansions of $\F$ and $\G$ in the $x$ variable:
\begin{eqnarray*}
\F(\xi,x;\hbar)=\sum_{q\in\Z^l}\F_q(\xi;\hbar)e^{i\la q,x\ra}
\\
\G(\xi,x;\hbar)=\sum_{q\in\Z^l}\G_q(\xi;\hbar)e^{i\la q,x\ra}
\end{eqnarray*}
where, $\forall q\in\Z^l$,
\begin{eqnarray*}
\F_q(\xi;\hbar)=(2\pi)^{-l/2}\int_{\R^l}\widehat{\F}_q(p;\hbar)e^{i\la p,\xi\ra}\,dp
\\
\G_q(\xi;\hbar)=(2\pi)^{-l/2}\int_{\R^l}\widehat{\G}_q(p;\hbar)e^{i\la p,\xi\ra}\,dp\,.
\end{eqnarray*}
If $\F_q(\xi;\hbar)\in\R$, and $\G_q(\xi;\hbar)\in\R,\;\forall q\in\Z^l$, then the Fourier expansion of $\{\F,\G\}_M $has purely imaginary Fourier coefficients, i.e.
$$
(\{\F,\G\}_M)_q(\xi;\hbar):= \int_{\R^l}\widehat{(\{\F,\G\}_M)}_q(p;\hbar)e^{i\la p,\xi\ra}\,dp\in i\R\,.
$$
\item[(4)]
Let $x\in\T^l\to\F(\xi,x;\hbar)\in\C$ and  $x\in\T^l\to\G(\xi,x;\hbar)\in\C$ belong to the space $\M$ of the functions with a definite parity (either even or odd) and let $J:\M\to\{-1,1\}$ be defined as in Section 2.3. Then
$$
J\{\F,\G\}_M = -(J\F)(J\G).
$$
\end{enumerate}
\end{theorem}
To prove the theorem we need the following
\begin{lemma}\label{La1}
Let $\F=\F(\xi,x;\hbar)\in C^{\infty}(\R^l\times\T^l\times[0,1]; \C)$.Then
\begin{enumerate}
\item[(i)]
$\F_q(\xi;\hbar)\in\R,\;\forall q\in\Z^l,\,\forall \xi\in\R^l$ if and only if 
$$
\overline{\widehat{\F}_q(p,\hbar)} = \widehat{\F}_q(-p,\hbar),\quad \forall q\in\Z^l,\;\forall p\in\R^l\,.
$$
\item[(ii)]
$\F_q(\xi;\hbar)\in i\R,\;\forall q\in\Z^l,\,\forall \xi\in\R^l$ if and only if 
$$
\overline{\widehat{\F}_q(p,\hbar)} = -\widehat{\F}_q(-p,\hbar),\quad \forall q\in\Z^l,\;\forall p\in\R^l\,.
$$
\end{enumerate}
\end{lemma}
{\it Proof of Lemma \ref{La1}.}
We prove only (i) because the proof of (ii) is analogous. If $\F_q(\xi;\hbar)\in\R,\;\forall q\in\Z^l,\,\forall \xi\in\R^l$, then
$$
\overline{\widehat{\F}_q(p,\hbar)} = (2\pi)^{-l/2}\int_{\R^l}\F_q(\xi,\hbar)e^{i\la p,\xi\ra}\,d\xi = (2\pi)^{-l/2}\int_{\R^l}\F_q(\xi,\hbar)e^{-i\la -p,\xi\ra}\,d\xi = \widehat{\F}_q(-p,\hbar)\,.
$$
Conversely, let $\overline{\widehat{\F}_q(p,\hbar)} = \widehat{\F}_q(-p,\hbar),\quad \forall q\in\Z^l,\;\forall p\in\R^l$.Then
\begin{eqnarray*}
\overline{\F_q(\xi;\hbar)} = (2\pi)^{-l/2}\int_{\R^l}\overline{\widehat{\F}_q(p,\hbar)}e^{-i\la p,\xi\ra}\,dp 
= (2\pi)^{-l/2}\int_{\R^l}\widehat{\F}_q(-p,\hbar)e^{i\la -p,\xi\ra}\,dp
\\
= (2\pi)^{-l/2}\int_{\R^l}\widehat{\F}_q(p,\hbar)e^{i\la p,\xi\ra}\,dp = \F_q(\xi;\hbar),
\end{eqnarray*}
where to obtain the third equality we have performed the change of variables $p\to -p$ in the integral. Hence $\F_q(\xi;\hbar)\in \R,\;\forall q\in\Z^l,\,\forall \xi\in\R^l$ and this completes the prooof of the lemma.
\pn
{\it Proof of Theorem \ref{Ta1}.} 
\begin{itemize}
\item[(1)]
If $\F$ and $\G$ do not depend on $x$, then $\F_q(\xi,\hbar)=\G_q(\xi,\hbar)=0,\;\forall q\neq 0, \forall \xi\in \R^l$. Therefore all the terms of the expansion in (\ref{M2}) with $q'\neq 0$ vanish. Then $\forall q\in\Z^l$
$$
\widehat{(\{\F,\G\}_M)}_q(p,\hbar) = \frac{2}{\hbar} \int_{\R^l}\widehat{\F}_{q}(p-p',\hbar)\widehat{\G}_{0}(p',\hbar)\sin( \frac{2}{\hbar}\la p',q\ra)\,dp'
$$
vanishes both for $q\neq 0$ and for $q=0$, whence $\{\F,\G\}_M\equiv 0$ by (\ref{M1}).
\item[(2)]
If $\G(\xi,x;\hbar)=\la\om,\xi\ra$, then by (\ref{M2})
\begin{eqnarray*}
\widehat{(\{\F,\G\}_M)}_q(p,\hbar) = \frac{2}{\hbar} \int_{\R^l}\widehat{\F}_{q}(p-p',\hbar)\widehat{\G}_{0}(p',\hbar)\sin( \frac{2}{\hbar}\la p',q\ra)\,dp' 
\\
= \frac{2}{\hbar} \int_{\R^l}\widehat{\F}_{q}(p-p',\hbar)\la\om, i\delta'(p)\ra\sin( \frac{2}{\hbar}\la p',q\ra)\,dp'
\\
=\frac{2i}{\hbar}\sum_{j=1}^l\om_j\frac{\partial}{\partial p_j}[\widehat{\F}_{q}(p-p',\hbar)\sin( \frac{2}{\hbar}\la p',q\ra)]|_{p'=0}
\\
= -i\sum_{j=1}^l\om_jq_j\widehat{\F}_{q}(p,\hbar) = -\widehat{\F}_{q}(p,\hbar)\la\om,iq\ra,
\end{eqnarray*}
where the Fourier transform $\widehat{\G}_{0}(p',\hbar)$ of $\G_0(\xi,\hbar) = \la\om,\xi\ra$ exists in the distributional sense, and is given by $i\delta'(p')$, where $\delta'(p')$ denotes the distributional derivative of the $\delta$-function:
$$
\int_{\R^l}\delta'(p')f(p')\,dp' = (\nabla_{p'}f)(0) = \sum_{j=1}^l\frac{\partial f}{\partial p_j'}|_{p'=0}\,,\qquad\forall  f\in\S(\R^l).
$$
Here $\S(\R^l)$ denotes the Schwartz space. Then by (\ref{M1})
\begin{eqnarray*}
\{\F,\G\}_M(\xi,x;\hbar) = -\int_{\R^l}\sum_{q\in\Z^l}\la\om,iq\ra\widehat{\F}_{q}(p,\hbar)
e^{i(\la p,\xi\ra\la + \la q,x\ra)}\,dp 
\\
= - \sum_{q\in\Z^l}\la\om,iq\ra\F_{q}(\xi,\hbar)
e^{i \la q,x\ra} = -\la\om,\nabla_x\F(\xi,x)\ra.
\end{eqnarray*}
\item[(3)] 
By Lemma \ref{La1} (i) we have $\overline{\widehat{\F}_q(p,\hbar)} = \widehat{\F}_q(-p,\hbar)$ and $\overline{\widehat{\G}_q(p,\hbar)} = \widehat{\G}_q(-p,\hbar),\; \forall q\in\Z^l,\;\forall p\in\R^l$. Then, from (\ref{M2}) we obtain
$$
\overline{\widehat{(\{\F,\G\}_M)}}_q(p,\hbar) = \frac{2}{\hbar} \int_{\R^l}\sum_{q'\in\Z^l}\widehat{\F}_{q-q'}(-p+p',\hbar)\widehat{\G}_{q'}(-p',\hbar)\sin[ \frac{2}{\hbar}(\la p',q\ra-\la p,q'\ra)]\,dp'
$$
whence, performing the change of variables $p'\to -p'$ in the integral,
\begin{eqnarray*}
\overline{\widehat{(\{\F,\G\}_M)}}_q(p,\hbar) = \frac{2}{\hbar} \int_{\R^l}\sum_{q'\in\Z^l}\widehat{\F}_{q-q'}(-p-p',\hbar)\widehat{\G}_{q'}(p',\hbar)\sin[ \frac{2}{\hbar}(-\la p',q\ra+\la -p,q'\ra)]\,dp'
\\
= -\frac{2}{\hbar} \int_{\R^l}\sum_{q'\in\Z^l}\widehat{\F}_{q-q'}(-p-p',\hbar)\widehat{\G}_{q'}(p',\hbar)\sin[ \frac{2}{\hbar}(\la p',q\ra-\la -p,q'\ra)]\,dp'
\\
= - \widehat{(\{\F,\G\}_M)}_q(-p,\hbar)\,,\quad\forall q\in\Z^l,\;\forall p\in\R^l.
\end{eqnarray*}
Then, by Lemma \ref{La1} (ii), $\widehat{(\{\F,\G\}_M)}(\xi,\hbar)\in i\R,\;\forall q\in\Z^l,\;\forall \xi\in\R^l$.
\item[(4)] 
First of all recall that $J\F=\pm 1$ if and only if $\F_q(\xi,\hbar)=\pm\F_{-q}(\xi,\hbar),\;\forall q\in\Z^l,\;\forall (\xi,\hbar)\in\R^l\times[0,1]$. Then by (\ref{M2}) we have
\begin{eqnarray*}
\widehat{(\{\F,\G\}_M)}_{-q}(p,\hbar) = \frac{2}{\hbar} \int_{\R^l}\sum_{q'\in\Z^l}\widehat{\F}_{-q-q'}(p-p',\hbar)\widehat{\G}_{q'}(p',\hbar)\sin[ \frac{2}{\hbar}(-\la p',q\ra-\la p,q'\ra)]\,dp'
\\
 = \frac{2}{\hbar} \int_{\R^l}\sum_{q'\in\Z^l}\widehat{\F}_{-q+q'}(p-p',\hbar)\widehat{\G}_{-q'}(p',\hbar)\sin[ \frac{2}{\hbar}(-\la p',q\ra+\la p,q'\ra)]\,dp'
\\
 = - \frac{2}{\hbar} \int_{\R^l}\sum_{q'\in\Z^l}\widehat{\F}_{-q+q'}(p-p',\hbar)\widehat{\G}_{-q'}(p',\hbar)\sin[ \frac{2}{\hbar}(\la p',q\ra-\la p,q'\ra)]\,dp',
\end{eqnarray*}
where in the second equality we have performed the change of variables $q'\to -q'$. Assume first that $J\F=J\G$; then $\F_{-q}\G_{-q}\equiv\F_{q}\G_{q}$ and $\widehat{\F}_{-q}\widehat{\G}_{-q}\equiv\widehat{\F}_{q}\widehat{\G}_{q},\;\forall q, q'\in\Z^l$. Thus,
\begin{eqnarray*}
\widehat{(\{\F,\G\}_M)}_{-q}(p,\hbar) 
= - \frac{2}{\hbar} \int_{\R^l}\sum_{q'\in\Z^l}\widehat{\F}_{q-q'}(p-p',\hbar)\widehat{\G}_{q'}(p',\hbar)\sin[ \frac{2}{\hbar}(\la p',q\ra-\la p,q'\ra)]\,dp'
\\
= -\widehat{(\{\F,\G\}_M)}_{q}(p,\hbar) ),
\end{eqnarray*}
whence
$$
(\{\F,\G\}_M)_{-q}(\xi,\hbar)  = - (\{\F,\G\}_M)_{q}(\xi,\hbar),\quad\forall q\in\Z^l,\;\forall (\xi,\hbar)\in\R^l\times[0,1]
$$
and $J\{\F,\G\}_M = -1= -(J\F)(J\G)$. In a similar way we obtain $J\{\F,\G\}_M = 1$ if $J\F=-J\G$, and this completes the proof of the theorem.
\end{itemize}
\subsection{The Weyl quantization}
Let us sum up the canonical (Weyl)  quantization procedure for functions (classical observables) 
defined on the phase space $\R^l\times\T^l$.  For more detail the reader is referred to \cite{GP}.
\par
 Let  $\A(\xi,x,\hbar):\R^l\times\T^l\times [0,1]\to\C$ be a family of smooth phase-space functions indexed by $\hbar$ fulfilling the assumptions of Theorem \ref{Ta1}, 
written under its Fourier representation
$$
\A(\xi,x,\hbar)=\int_{\R^l}\sum_{q\in\Z^l}\widehat{\A}_q(p;\hbar)e^{i(\la p,\xi\ra +\la q,x\ra)}\,dp
$$
where, as in Section 1:
\begin{eqnarray*}
&&
\A(\xi,x,\hbar)=\sum_{q\in\Z^l}\,\A_q(\xi,\hbar) e^{i\la q,x\ra}, \qquad \A_q(\xi,\hbar):=(2\pi)^{-l/2}\int_{\T^l}\,\A(\xi,x;\hbar)e^{-i\la q,x\ra}\,dx
\\
&&
\widehat{\A}_q(p;\hbar)=(2\pi)^{-l/2}\int_{\R^l}\,\A_q(\xi;\hbar)e^{-i\la p,\xi\ra}\,dx
\end{eqnarray*}
Then the (Weyl) quantization of  $\A(\xi,x;\hbar)$  is the operator acting on $L^2(\T^l)$, defined by:
\be
\label{(A1)}
(A(\hbar)f)(x):=
\int_{\R^l}\sum_{q\in\Z^l}\widehat{\A}_q(p;\hbar)e^{i(\la q,x\ra+\la p,q\ra\hbar/2 )}f(x+p\hbar)\,dp,\;  f\in L^2(\T^l).
\ee
\begin{remark}\label{Ra1}
\begin{enumerate}
{\rm \item[(a)]
If $\A$ does not depend on $\xi$,  $\A(\xi,x,\hbar)=\A(x,\hbar)$, (A.1) reduces to the standard {\it multiplicative} action:
\begin{eqnarray*}
&&
(A(\hbar)f)(x)
=\int_{\R^l}\sum_{q\in\Z^l}\A_q(\hbar)\delta(p)e^{i(\la q,x\ra+\la p,q\ra \hbar/2)}f(x+\hbar p)\,dp 
\\
&&
=\sum_{q\in\Z^l}\A_q(\hbar)e^{i\la q,x\ra}f(x)=\A(x,\hbar)f(x)
\end{eqnarray*}
\item[(b)]
If $\A$ does not depend on $x$, then $\widehat{\A}_q=0, q\neq 0$; 
thus $\widehat{\A}_0=\widehat{\A}(p,\hbar)$ and the standard (pseudo) {differential action} is recovered:
\begin{eqnarray*}
(A(\hbar)f)(x)&=&\ds\int_{\R^l}\widehat{\A}(p,\hbar)f(x+\hbar p)\,dp 
=\int_{\R^l}\sum_{q\in\Z^l}\,\widehat{\A}(p,\hbar)f_qe^{i\la q,x+\hbar p\ra}\,dp
\\
& =& \sum_{q\in\Z^l}f_q\A(q\hbar,\hbar)e^{i\la q,x\ra}
 = (\A(-i\hbar\nabla_x,\hbar)f)(x),
\end{eqnarray*}
whence the {formula} yielding all the eigenvalues of $A$:
\be
\label{A2}
\lambda_n(\hbar)=\la e_n,Ae_n\ra =\A(n\hbar,\hbar).
\ee
where $\{e_n:n\in\Na\}$ is the set of the Hermite functions in $L^2(\R^l)$.
\item[(c)]
Let $\V(t,x;\hbar)$ be a complex-valued,  smooth function defined $\R\times\T^l\times[0,1]$ vanishing exponentially fast as $|t|\to\infty$ uniformly w.r.t. $(x,\hbar)\in\T^l\times[0,1]$,  with Fourier expansion 
\be
\label{A3}
\V(t,x;\hbar)=\int_{\R}\sum_{q\in\Z^l}\widehat{\V}_q(p;\hbar)e^{i(\la p,\xi\ra +\la q,x\ra)}\,dp
\ee
where, as in Section 1:
\begin{eqnarray*}
&&
\V(t,x,\hbar)=\sum_{q\in\Z^l}\,\V_q(t,\hbar) e^{i\la q,x\ra}, \qquad \V_q(t,\hbar):=(2\pi)^{-l/2}\int_{\T^l}\,\V(t,x;\hbar)e^{-i\la q,x\ra}\,dx
\\
&&
\widehat{\V}_q(p;\hbar)=(2\pi)^{-l/2}\int_{\R}\,\V_q(t;\hbar)e^{-i\la p,t\ra}\,dt
\end{eqnarray*}
and  let the smooth function $\V_\om(\xi,x;\hbar): \R^l\times\T^l\times [0,1]\to\C$ be defined as follows:
$$
\V_\om(\xi,x;\hbar):=\left.\V(t,x,\hbar)\right|_{t=\L_\om(\xi)}=\V(\la\om,\xi\ra,x;\hbar).
$$
Then we have:
$$
\V_\om(\xi,x;\hbar)=\int_\R\,\sum_{q\in\Z^l}\,\widehat{\V}_q(p,\hbar)e^{i(\la q,x\ra+p\L_\om(\xi))}\,dp
$$
and (A.1) clearly becomes:
\be
\label{A4}
(V_{\om}(\hbar)f)(x)
=\int_{\R}\sum_{q\in\Z^l}\widehat{\V}_q(p;\hbar)e^{i(\la q,x\ra+p\la \om,q\ra \hbar/2 )}f(x+p\hbar \om)\,dp
\ee
\item[(d)]
 Let
 \be
\label{A5}
\|\V_{\om}\|_{\rho}:=\sup_{\hbar\in[0,1]}\,\sum_{q\in\Z^l}\,e^{\rho |q|}\,\int_{\R}\,e^{\rho |p|}\,|\widehat{\V}_q(p,\hbar)|\,dp<+\infty, \quad \rho\geq 0.
\ee
and remark that
\begin{eqnarray*}
\|\V_{\om}\|_{L^1}:=\sup_{\hbar\in[0,1]}\,\sum_{q\in\Z^l}\,\int_{\R}\,|\widehat{\V}_q(p,\hbar)|\,dp \leq  \|\V_{\om}\|_\rho.
\end{eqnarray*}
\vskip 4pt\noindent
Then $V_{\om}(\hbar)$ is a  bounded operator in $L^2(\T^l)$, uniformly with respect to $\hbar\in [0,1]$, namely:
\be
\label{A6}
\sup_{\hbar\in[0,1]} \| V_{\om}(\hbar)\|_{L^2\to L^2} \leq \|\V_{\om}\|_{L^1}\leq \|\V_{\om}\|_\rho
\ee
because
\begin{eqnarray*}
&&
\| V_{\om}(\hbar)f\|_{L^2}\leq \sum_{q\in\Z^l}\int_{\R}\,|\widehat{\V}_q(p,\hbar)|\,dp \,\|f\|_{L^2}
\leq \|\V_{\om}\|_{L^1}\,\|f\|_{L^2}.
\end{eqnarray*}
\item[(e)]
If the symbol $\V$ is real valued, then its Weyl quantization $V(\hbar)$ is a clearly symmetric operator in $L^2(\T^l)$; if in addition condition (A.5) holds its boundedness  entails its self-adjointness.}
\end{enumerate}
\end{remark}
\end{appendix}

\end{document}